\documentclass[twocolumn, traditabstract]{aa} %%% FOR ONE COLUMN
\usepackage{graphicx}
\usepackage{amsmath}
\usepackage{bbm}
\usepackage{layouts}
\usepackage{color}
    \definecolor{Blue}{rgb}{0.0,0.0,1.0}
    \definecolor{Red}{rgb}{1.0,0.0,0.0}
    \definecolor{Green}{rgb}{0.0,1.0,0.0}
\begin{document}
%-----------------------------------------------------------------------
\title{The perihelion of Mercury advance \\calculated in Newton's theory}
%
%   \subtitle{}
%-----------------------------------------------------------------------
\author{       M.A. Abramowicz\inst{1, 2, 3}
%-----------------------------------------------------------------------
%-----------------------------------------------------------------------
}
%-----------------------------------------------------------------------
\institute{Klinika Chirurgii Gastroenterologicznej i Transplantologii Centralnego Szpitala
Klinicznego MSW, Warszawa, Poland
%-----------------------------------------------------------------------
\and          Physics Department, Gothenburg University,
               SE-412-96 G{\"o}teborg, Sweden
                 \\ \email{marek.abramowicz@physics.gu.se}
%-----------------------------------------------------------------------
\and           Copernicus Astronomical Center, ul. Bartycka 18, PL-00-716
               Warszawa, Poland
               }
%-----------------------------------------------------------------------
   \date{Received ????; accepted ????}
%-----------------------------------------------------------------------
%-----------------------------------------------------------------------
\abstract {Three radii are associated with a circle: the ``geodesic radius'' $r_*$ which is
the distance from circle's center to its perimeter, the ``circumferential radius'' ${\tilde
r}$ which is the length of the perimeter divided by $2\pi$ and the ``curvature radius''
${\cal R}$ which is circle's curvature radius in the Frenet sense.  In the flat Euclidean
geometry it is $r_* = {\tilde r} = {\cal R}$, but in a curved space these three radii are
different. I show that although Newton's dynamics uses Euclidean geometry, its equations
that describe circular motion in spherical gravity always {\it unambiguously} refer to one
{\it particular} radius of the three --- geodesic, circumferential, or curvature. For
example, the gravitational force is given by $F = -GMm/{\tilde r}^2$, and the centrifugal
force by $mv^2/{\cal R}$. Building on this, I derive a Newtonian formula for the perihelion
of Mercury advance.}
%-----------------------------------------------------------------------
%-----------------------------------------------------------------------
%-----------------------------------------------------------------------
\authorrunning{Abramowicz }\titlerunning{Perihelion of Mercury advance}
%-----------------------------------------------------------------------
\maketitle
%-----------------------------------------------------------------------
%
%#######################################################################
%-----------------------------------------------------------------------
\section{Introduction}
\begin{figure}[b!]
\begin{center}
\includegraphics[width=0.435\textwidth]{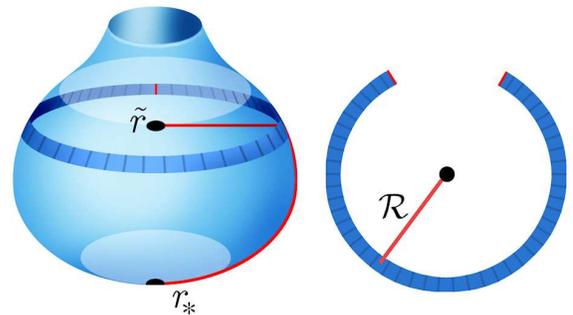}
\caption{Geodesic $r_*$, circumferential ${\tilde r}$, and curvature ${\cal R}$ radii of a
circle on a curved 2-D surface. The curvature radius may be calculated from the other two,
${\cal R} = {\cal R}(r_*, {\tilde r})$.}
\end{center}
\label{Figure}
\end{figure}

%-----------------------------------------------------------------------
%#######################################################################
%
Newton's theory of gravity was formulated in a flat, Euclidean 3-D space but its basic laws,
i.e. the Poisson equation and the equation of motion,
%-----------------------------------------------------------------------
\begin{eqnarray}
\hskip 2truecm
g^{ik} \nabla_i \nabla_k \Phi &=& -4\pi G \rho,
\label{Poisson-equation}
\\
F_i &=& m a_i,
\label{equation-of-motion}
\end{eqnarray}
%-----------------------------------------------------------------------
make a perfect sense in the 3-D space with an arbitrary geometry $g_{ik}$. Here, I will show
that some of the geometrical concepts in a curved space {\it naturally} pop-up from the
Newtonian dynamics. In particular, Newton's dynamics {\it knows} about differences between
the three radii of a circle: the geodesic radius $r_*$, the circumferential radius ${\tilde
r}$, and the curvature radius ${\cal R}$ (shown in Figure~\ref{Figure}; in the next Section,
we give their formal definitions). Therefore, {\it these radii may be measured by Newtonian
dynamical experiments}.
%ooooooooooooooooooooooooooooooooooooooooooooooooooooooooooooooooooooooo
%-----------------------------------------------------------------------
%-----------------------------------------------------------------------
%ooooooooooooooooooooooooooooooooooooooooooooooooooooooooooooooooooooooo
In the flat, Euclidean,  space the three radii are equal, but in a space with a non-zero
Gaussian curvature ${\cal G}$, they are different\footnote{The Gaussian curvature at a
particular point of a 2-D surface is given by its two principal curvature radii, ${\cal G} =
1/(R_1 R_2)$. For a sphere with radius $R$ it is ${\cal G} = 1/(R^2)$, for a cylinder it is
${\cal G} = 0$. According to Gauss's {\it Theorema Egregium}, the curvature ${\cal G}$ may
be determined by intrinsic geometry of the surface, with no reference to the external 3-D
Euclidean space. Note that the dimension of the Gaussian curvature is $1/({\rm
lenght})^2$.}. The Gaussian curvature may be calculated from the explicitly known relation
${\cal G} = {\cal G}(r_*, {\tilde r})$, which means that it can be {\it measured} from
Newtonian experiments. Then, {\it the perihelion of the Mercury advance $\Delta \phi$ may be
explained in Newton's theory} because, as we will show later in this article, Newton's
theory predicts,
%-----------------------------------------------------------------------
\begin{equation}
\Delta \phi = - \pi \left( \frac{d\,{\cal G}}{\,d\, r_*} \right)^2 r_*^6.
\label{perihelion-explained}
\end{equation}
%-----------------------------------------------------------------------
%
%#######################################################################
%-----------------------------------------------------------------------
\section{The three radii of a circle}
%-----------------------------------------------------------------------
%#######################################################################
Consider a two dimensional curved surface with the metric
%-----------------------------------------------------------------------
\begin{equation}
ds^2 = dr_*^2 + [{\tilde r(r_*)}]^2d\phi^2.
\label{metric}
\end{equation}
%-----------------------------------------------------------------------
A circle is defined by the condition $r_*\,=\,\,$const. The geodesic radius of the circle
equals $r_*$. This follows from
%-----------------------------------------------------------------------
\begin{equation}
({\rm geodesic~radius}) \equiv \int_0^{r_*} ds_{\vert \phi = {\rm const}} = \int_0^{r_*}
dr_* = r_*.
\label{definition-r-star}
\end{equation}
%-----------------------------------------------------------------------
The circumferential radius of the circle equals ${\tilde r}$. This follows from
%-----------------------------------------------------------------------
\begin{eqnarray}
\hskip0.5truecm
({\rm circumferential~radius}) \equiv \frac{1}{2\pi}\int_0^{2\pi} ds_{\vert r_* = {\rm
const}}&& \nonumber \\
~~~~~~~~~~~~ = \frac{1}{2\pi} \int_0^{2 \pi}{\tilde r} d\phi = {\tilde r}.&&
\label{definition-r-tilde}
\end{eqnarray}
%-----------------------------------------------------------------------
%
A unit tangent vector to a circle is defined by
%-----------------------------------------------------------------------
\begin{equation}
\tau^i = \frac{1}{{\tilde r}}\delta^i_{\,\phi},
\label{unit-vector}
\end{equation}
%-----------------------------------------------------------------------
where $\delta^i_{\,\phi}$ is the Kronecker delta. The curvature radius ${\cal R}$ of a
circle is defined by,
%-----------------------------------------------------------------------
\begin{equation}
({\rm curvature~radius}) \equiv \left[ \frac{d\tau^i}{ds}\,\frac{d\tau_i}{ds}\right]^{-1/2}
= {\cal R} .
\label{curvature-radius}
\end{equation}
%-----------------------------------------------------------------------
This definition follows from the Frenet formula,
%-----------------------------------------------------------------------
\begin{equation}
\frac{d\tau^i}{ds} = - \frac{1}{{\cal R}} \lambda^i,
\label{curvature-radius}
\end{equation}
%-----------------------------------------------------------------------
where $\lambda^i$ is a unit vector normal to the circle.

We will use later two useful formulae for the curvature of the circle ${\cal K} = 1/{\cal
R}$, and for the Gaussian curvature ${\cal G}$ of the surface with the metric
(\ref{metric}),
%-----------------------------------------------------------------------
\begin{eqnarray}
\hskip2truecm
{\cal K} &=& +\, \frac{1}{{\tilde r}}\left( \frac{d{\tilde r}}{\,dr_*} \right),
\label{radius-curvature-useful} \\
{\cal G} &=& -\, \frac{1}{{\tilde r}} \left( \frac{d^{2}{\tilde r}}{d r_*^2}\right).
\label{Gauss-curvarure-useful}
\end{eqnarray}
%-----------------------------------------------------------------------
%
Formula (\ref{radius-curvature-useful}) follows from (\ref{curvature-radius}). For
derivation of (\ref{Gauss-curvarure-useful}) see e.g. \cite{Synge-1978}, Section 3.4.
%#######################################################################
%-----------------------------------------------------------------------
\section{Equations of motion}
%-----------------------------------------------------------------------
%#######################################################################
%
Let us consider a curve in space given by a parametric equation,
%-----------------------------------------------------------------------
\begin{equation}
x^i = x^i(s),
\label{parametric-equation}
\end{equation}
%-----------------------------------------------------------------------
where $x^i$ are coordinates in space, and $s$ is the length along the curve. If a body moves
along this curve, its velocity equals,
%-----------------------------------------------------------------------
\begin{equation}
v^i = \frac{dx^i}{dt} = \frac{ds}{dt} \frac{dx^i}{ds} = v \tau^i.
\label{velocity}
\end{equation}
%-----------------------------------------------------------------------
Here $v =ds/dt$ is the speed of the body and $\tau^i = dx^i/ds$ is a unit vector tangent to
the curve (\ref{parametric-equation}), i.e. the direction of motion. The acceleration may be
calculated as follow,
%-----------------------------------------------------------------------
\begin{equation}
a^i = \frac{dv^i}{dt} = \frac{ds}{dt} \frac{d(v \tau^i)}{ds} = v^2 \left(
\frac{d\tau^i}{ds}\right) + \tau^i v\frac{dv}{ds}.
\label{acceleration}
\end{equation}
%-----------------------------------------------------------------------
Assuming a particular of circular motion with constant velocity, $v = $ const, and applying
(\ref{curvature-radius}) to calculate the term in brackets, we arrive at
%-----------------------------------------------------------------------
\begin{equation}
a^i = v^2\frac{1}{{\cal R}} \lambda^i,
\label{acceleration}
\end{equation}
%-----------------------------------------------------------------------
which is the well known formula for the centrifugal acceleration.

Consider now a circular motion around a spherically symmetric center of gravity. The
Newtonian equation of motion, $F^i = m a^i$, takes the form,
%-----------------------------------------------------------------------
\begin{equation}
- \nabla^i \Phi = v^2\frac{1}{{\cal R}} \lambda^i,
\label{equation-motion}
\end{equation}
%-----------------------------------------------------------------------
where $F^i = - m \nabla^i \Phi$ is the gravitational force, and $\Phi$ is the gravitational
potential. Three quantities characterize motion at a particular circular orbit: the angular
velocity $\Omega$, the angular speed $v$, and the specific angular momentum ${\cal L}$. They
are related by,
%-----------------------------------------------------------------------
\begin{eqnarray}
\hskip 3.5truecm
v &=& {\tilde r} \Omega,
\label{velocity-speed-momentum-1}
\\
%........................................................................
{\cal L} &=& {\tilde r} v = {\tilde r}^2 \Omega.
\label{velocity-speed-momentum-2}
\end{eqnarray}
%-----------------------------------------------------------------------
Using (\ref{velocity-speed-momentum-2}), and multiplying its left and right side by
$\lambda_i$, we transform the equation of motion (\ref{equation-motion}) into a form which
will be convenient later,
%-----------------------------------------------------------------------
\begin{equation}
\lambda_i \nabla^i \Phi = \frac{{\cal L}^2}{{\tilde r}^2 {\cal R}}.
\label{equation-motion-1}
\end{equation}
%-----------------------------------------------------------------------
Let us remind that $\lambda^i$ is a unit, {\it outside pointing}, vector. Here ''outside''
has the absolute meaning --- outside the center, in the direction to infinity. We will
calculate the left-hand side of this equation in the next Section.

%-----------------------------------------------------------------------
\section{Newton's gravity and Kepler's law}
%-----------------------------------------------------------------------
In an empty space, the gravitational potential $\Phi$ obeys the Laplace equation,
%-----------------------------------------------------------------------
\begin{equation}
\nabla_i (\nabla^i \Phi) = 0.
\label{Laplace}
\end{equation}
%-----------------------------------------------------------------------
Let us integrate (\ref{Laplace}) over the volume $\mathbbm{V}$ that is contained between two
spheres, concentric with the gravity center, with sphere $\mathbbm{S}_1$ being inside sphere
$\mathbbm{S}_2$. We transform the volume integral into the surface integral, using the Gauss
theorem
%-----------------------------------------------------------------------
\begin{eqnarray}
0 &=& \int_{\mathbbm{V}}\nabla_i (\nabla^i \Phi) d\mathbbm{V} \nonumber \\
&=& \int_{\mathbbm{S}_1}(\nabla^i \Phi) N^{(1)}_i d\mathbbm{S} +
\int_{\mathbbm{S}_2}(\nabla^i \Phi) N^{(2)}_i d\mathbbm{S}.
\label{Gauss}
\end{eqnarray}
%-----------------------------------------------------------------------
The oriented surface elements on the spherical surfaces $\mathbbm{S}_1$ and $\mathbbm{S}_2$
may be written, respectively, as
%-----------------------------------------------------------------------
\begin{equation}
N^{(1)}_i d\mathbbm{S} = - \lambda_i d\mathbbm{S}, ~~N^{(2)}_i d\mathbbm{S} = + \lambda_i
d\mathbbm{S},
\label{surface-elements}
\end{equation}
%-----------------------------------------------------------------------
and therefore,
%-----------------------------------------------------------------------
\begin{equation}
\int_{\mathbbm{S}_1}(\nabla^i \Phi) \lambda_i d\mathbbm{S} = \int_{\mathbbm{S}_2}(\nabla^i
\Phi) \lambda_i d\mathbbm{S}.
\label{two-spheres}
\end{equation}
%-----------------------------------------------------------------------
This means that the value integral is the same, say $S_0$, for all spheres around the
gravity center. In addition, because of the spherical symmetry of the potential, the
quantity $(\nabla^i \Phi) \lambda_i$ is constant over the sphere of integration. Thus,
%-----------------------------------------------------------------------
\begin{eqnarray}
S_0 &=&  (\nabla^i \Phi) \lambda_i \int_{\mathbbm{S}} d\mathbbm{S} = 4\pi {\tilde
r}^2\, (\nabla^i \Phi) \lambda_i, ~~ {\rm and} \nonumber \\
(\nabla^i \Phi) \lambda_i &=& \frac{S_0}{4\pi {\tilde r}^2} = \frac{GM}{{\tilde r}^2},
\label{one-sphere}
\end{eqnarray}
%-----------------------------------------------------------------------
Combining (\ref{one-sphere}) with (\ref{equation-motion-1}), we may finally write,
%-----------------------------------------------------------------------
\begin{equation}
{\cal L}^2 = GM{\cal R}.
\label{Kepler-angular-momentum}
\end{equation}
%-----------------------------------------------------------------------
This is the Kepler Third Law. Using natural units for radius and frequency,
%-----------------------------------------------------------------------
\begin{eqnarray}
\hskip 2truecm
R_G &=& \frac{GM}{c^2},
\label{gravitational-radius} \\
\Omega_G &=& \frac{c^3}{GM},
\label{gravitational-frequency}
\end{eqnarray}
%-----------------------------------------------------------------------
we may write the formula for the Keplerian angular velocity as,
%-----------------------------------------------------------------------
\begin{equation}
\left(\frac{{\Omega}}{\,\Omega_G}\right)^2 = R_G^3 \left(\frac{{\cal R}}{{\tilde
r}^4}\right).
\label{Kepler-angular-velocity}
\end{equation}
%-----------------------------------------------------------------------
%
%#######################################################################
%-----------------------------------------------------------------------
\section{Epicyclic oscillations, the perihelion advance }
%-----------------------------------------------------------------------
%#######################################################################
%
Suppose that we slightly perturb a test-body on a circular orbit. This means that its
angular momentum will not correspond to the Keplerian one, ${\cal L}^2$, given by
(\ref{Kepler-angular-momentum}), but will be slightly different ${\cal L}^2 + \delta{\cal
L}^2$. There will be also a small radial motion with velocity ${\dot{(\delta r_*)}}$ and
acceleration ${\ddot{(\delta r_*)}}$. From (\ref{equation-motion-1}) it follows that
%-----------------------------------------------------------------------
\begin{equation}
\frac{GM}{{\tilde r}^2} - \frac{{\cal L}^2 + \delta{\cal L}^2}{{\tilde r}^2 {\cal R}} =
{\ddot{(\delta r_*)}}.
\label{perturbation}
\end{equation}
%-----------------------------------------------------------------------
Keeping the first order term in equation (\ref{perturbation}), and using
%-----------------------------------------------------------------------
\begin{equation}
\delta{\cal L}^2 = \frac{d{\cal L}^2}{d r_*}\,(\delta r_*),
\label{perturbation-delta-L}
\end{equation}
%-----------------------------------------------------------------------
we arrive at the simple harmonic oscillator equation,
%-----------------------------------------------------------------------
\begin{equation}
\omega^2 (\delta r_*) + {\ddot{(\delta r_*)}} = 0,
\label{simple-harmonic}
\end{equation}
%-----------------------------------------------------------------------
where $\omega$ is the radial epicyclic frequency,
%-----------------------------------------------------------------------
\begin{equation}
\omega^2 = \frac{1}{{\tilde r}^2 {\cal R}}\left(\frac{d{\cal L}^2}{d r_*}\right).
\label{epicyclic-frequency-1}
\end{equation}
%-----------------------------------------------------------------------
Using equations (\ref{Kepler-angular-momentum}), (\ref{gravitational-radius}) and
(\ref{gravitational-frequency}), we may write the expression for the epicyclic frequency in
the form,
%-----------------------------------------------------------------------
\begin{equation}
\left( \frac{\omega}{\Omega_G}\right)^2 = \left(\frac{d{\cal R}}{\,\,d r_*}\right) \frac
{R_G^3}{{\tilde r}^2 \,{\cal R}},
\label{epicyclic-frequency-2}
\end{equation}
%-----------------------------------------------------------------------
or comparing this with (\ref{Kepler-angular-velocity}),
%-----------------------------------------------------------------------
\begin{equation}
\left( \frac{\omega}{\Omega}\right)^2 \,=\, \left(\frac{d{\cal R}}{\,\,d r_*}\right) \frac
{{\tilde r}^2}{{\cal R}^2} \,=\, \left(\frac{d {\tilde r}}{\,\,d r_*} \right)^2 - {\tilde r}
\left( \frac{d^2 {\tilde r}}{d r_*^2} \right).
\label{epicyclic-frequency-3}
\end{equation}
%-----------------------------------------------------------------------
In a flat space, $r_*$$\,=\,$${\tilde r}$$\,=\,$${\cal R}$, and therefore
$\omega$$\,=\,$$\Omega$, which implies that the slightly non-circular orbit is a closed
curve, indeed an ellipse. In a curved space with ${\cal G}$$\,\not=\,$$0$, it is
$r_*$$\,\not=\,$${\tilde r}$$\,\not=\,$${\cal R}$, and consequently
$\omega$$\,\not=\,$$\Omega$. The slightly non-circular orbit would not be a closed curve. It
could be represented by a precessing ellipse, with two consecutive perihelia shifted by
%-----------------------------------------------------------------------
\begin{equation}
\Delta \phi = T(\Omega - \omega) = 2\pi(1 - \frac{\omega}{\Omega}),
\label{perihelion-shift-1}
\end{equation}
%-----------------------------------------------------------------------
where $T = 2\pi/\Omega$ is the orbital period. Let us consider a particular form of the
metric (\ref{metric}), with
%-----------------------------------------------------------------------
\begin{equation}
\left( g_{\phi \phi} \right)^{1/2} \equiv {\tilde r} = r_*\,\left[ 1 + \alpha \left(
\frac{r_*}{\mathbbm{R}} \right)
 \right],
\label{metric-example}
\end{equation}
%-----------------------------------------------------------------------
where $\mathbbm{R}$ and $\alpha$ are constant and $r_*/\mathbbm{R} \ll 1$. In this case one
has,
%-----------------------------------------------------------------------
\begin{eqnarray}
\left(\frac{d {\tilde r}}{\,\,d r_*} \right)^2 &=&
1 + 4\alpha \left( \frac{r_*}{\mathbbm{R}} \right), \\
\left( \frac{d^2{\tilde r}}{\,\,d r_*^2}\right) &=& \frac{1}{r_*} \left[ 2\alpha \left(
\frac{r_*}{\mathbbm{R}} \right) \right],
\\
{\tilde r}^2\,{\cal G} &=& - 2\alpha \left( \frac{r_*}{\mathbbm{R}} \right) .
\label{expansion-results}
\end{eqnarray}
%-----------------------------------------------------------------------

%#######################################################################
%-----------------------------------------------------------------------
\section{Surface of constant curvature}
%-----------------------------------------------------------------------
%#######################################################################
%
A 2-sphere with radius $\mathbbm{R}$ has a constant Gauss curvature ${\cal G} =
1/\mathbbm{R}^2$ and the metric,
%-----------------------------------------------------------------------
\begin{eqnarray}
ds^2 &=& dr_*^2 + {\tilde r}^2 d\phi^2 = dr_*^2 + \mathbbm{R}^2 \sin^2\left(
\frac{r_*}{\mathbbm{R}}\right) d\phi^2
\label{positive-curvature-metric} \\
&& {\tilde r} = r_* \left[ 1 - \frac{1}{6}\left( \frac{r_*}{\mathbbm{R}}  \right)^2 +
...\right].
\label{positive-curvature-metric-small}
\end{eqnarray}
%-----------------------------------------------------------------------
Similarly, a 2-space with constant negative curvature ${\cal G} = - 1/\mathbbm{R}^2$ has the
metric,
%-----------------------------------------------------------------------
\begin{eqnarray}
ds^2 &=& dr_*^2 + {\tilde r}^2 d\phi^2 = dr_*^2 + \mathbbm{R}^2 \sinh^2\left(
\frac{r_*}{\mathbbm{R}}\right) d\phi^2.
\label{negative-curvature-metric} \\
&& {\tilde r} = r_* \left[ 1 + \frac{1}{6}\left( \frac{r_*}{\mathbbm{R}}  \right)^2 +
...\right].
\label{negative-curvature-metric-small}
\end{eqnarray}
%-----------------------------------------------------------------------
This means that, in a space with small constant (positive or negative) curvature,
%-----------------------------------------------------------------------
\begin{equation}
\alpha = 0,
\label{positive-curvature-alpha-beta}
\end{equation}
%-----------------------------------------------------------------------
and therefore $\Delta \phi/2\pi = 0$. There is no perihelion precession in spaces with
constant (positive or negative) curvature.

This, together with equations (\ref{perihelion-shift-1}), (\ref{expansion-results}) derived
in the previous Section imply that the perihelion advance can be expressed by the derivative
of the Gaussian curvature of space,
%-----------------------------------------------------------------------
\begin{equation}
\frac{\Delta \phi}{2\pi} = -2\alpha^2 \left( \frac{r_*}{\mathbbm{R}} \right)^2
= -\frac{1}{2}\left( \frac{d\,{\cal G}}{\,d\, r_*} \right)^2 r_*^6 .
\label{final-equation}
\end{equation}
%-----------------------------------------------------------------------

%
%#######################################################################
%-----------------------------------------------------------------------
\section{Discussion and conclusions}
%-----------------------------------------------------------------------
%#######################################################################
%
Newton's theory of gravity was formulated in a flat, Euclidean 3-D space but its basic laws,
i.e. the Poisson equation and the equation of motion, make a perfect sense in the 3-D space
with an arbitrary geometry $g_{ik}$. In particular, Newtonian dynamics allows to {\it
measure} the circumferential ${\tilde r}$ and curvature ${\cal R}$ radii
of circular orbits by measuring the gravitational $F_G$ and centrifugal $F_C$ forces,
%------------------------------------------------------------------------
\begin{equation}
F_G = -\frac{GMm}{{\tilde r}^2}, ~~ F_C = \frac{m\,v^2}{{\cal R}}.
\label{gravitational-centrifugal}
\end{equation}
%------------------------------------------------------------------------
If ${\tilde r} \not= {\cal R}$, it is ${\cal G} \not= 0$. One may measure
the Gaussian curvature of space at different circular orbits and find
${\cal G} = {\cal G}(r_*)$ using formulae
(\ref{radius-curvature-useful}) and (\ref{Gauss-curvarure-useful}). The predicted
value of the perihelion advance is $\Delta \phi/2\pi =
-(1/2) (d{\cal G}/dr_*)^2 \, {\tilde r}^6$.

%#######################################################################
%-----------------------------------------------------------------------
\begin{acknowledgements}
Calculations presented in this article have been done during author's treatment at the
Gastroenterology and Trans\-plan\-tology Ward of the MSW Hospital in Warsaw, before and
after his surgery. The author thanks Dr Andrzej Otto who has performed the surgery. This
work has been supported by the Polish National Health Foundation (NFZ) and the NCN
UMO-2011/01/B/ST9/05439 grant.
\vskip0.01truecm \noindent
I would like to thank Maciej Wielgus for checking all calculations presented here.
\end{acknowledgements}
%-----------------------------------------------------------------------
%#######################################################################
%

%-----------------------------------------------------------------------

%-----------------------------------------------------------------------
\end{document}